\documentclass{elsart1p}

\usepackage{graphicx}
\usepackage{amssymb}
\usepackage{mathrsfs}
\usepackage{calrsfs}
\usepackage{txfonts}
\usepackage{pslatex}

\newcommand{\sqrtsnn}{\sqrt{s_{\mbox{\tiny{\it{NN}}}}}}
\newcommand{\AaAa}{$AA$}
\newcommand{\AuAu}{$AuAu$}
\newcommand{\CuCu}{$CuCu$}
\newcommand{\PbPb}{$PbPb$}

\newcommand{\dAu}{$dAu$}
\newcommand{\pp}{$pp$}

\newcommand\mqhat{{\mean{\hat{q}}}}
\newcommand\qhat{{\hat{q}}}

\newcommand{\ecrit}{\varepsilon_{\mbox{\tiny{\it crit}}}}

\newcommand{\IAAa}{I^{away}_{AA}}
\newcommand{\etg}{E_T^{\gamma}}
\newcommand{\etj}{E_T^{jet}}
\def\z{z_{_{\gamma h}}}

\newcommand{\fastjet}{{\sc FastJet}}

\def\mean#1{\ensuremath{\left<#1\right>}}

\begin{document}

\begin{frontmatter}

\title{Jet quenching in QCD matter: from RHIC to LHC}

\author{David d'Enterria}\ead{david.d'enterria@cern.ch}
\address{LNS, MIT, Cambridge, MA 02139-4307, USA}

\begin{abstract}
The current experimental and theoretical status of hadron and jet production at large transverse momentum 
in high-energy nucleus-nucleus collisions is summarised. The most important RHIC results 
are compared to theoretical parton energy  loss predictions providing direct information on the (thermo)dynamical 
properties of hot and dense QCD matter. Prospects for the LHC are also outlined.
\end{abstract}
\end{frontmatter}


\section{Introduction}
\label{sec:intro}

The physics programme of high-energy nucleus-nucleus (\AaAa) collisions is focused on the study 
of the fundamental theory of the strong interaction -- Quantum Chromo Dynamics (QCD) --
in extreme conditions of temperature, density and small parton momentum fraction (low-$x$) 
-- see e.g.~\cite{d'Enterria:2006su} for a recent review.
By colliding two heavy nuclei at ultrarelativistic energies one expects to form a hot and dense 
deconfined medium whose collective (colour) dynamics can be studied experimentally.
Lattice QCD calculations~\cite{latt} predict a new form of matter at energy densities (well)
above $\ecrit \approx$~1~GeV/fm$^3$ consisting of an extended volume of deconfined and
chirally-symmetric (bare-mass) quarks and gluons: the Quark Gluon Plasma (QGP). 

\begin{figure}[htbp]
\centering
\includegraphics[width=0.49\linewidth,height=6.5cm,clip]{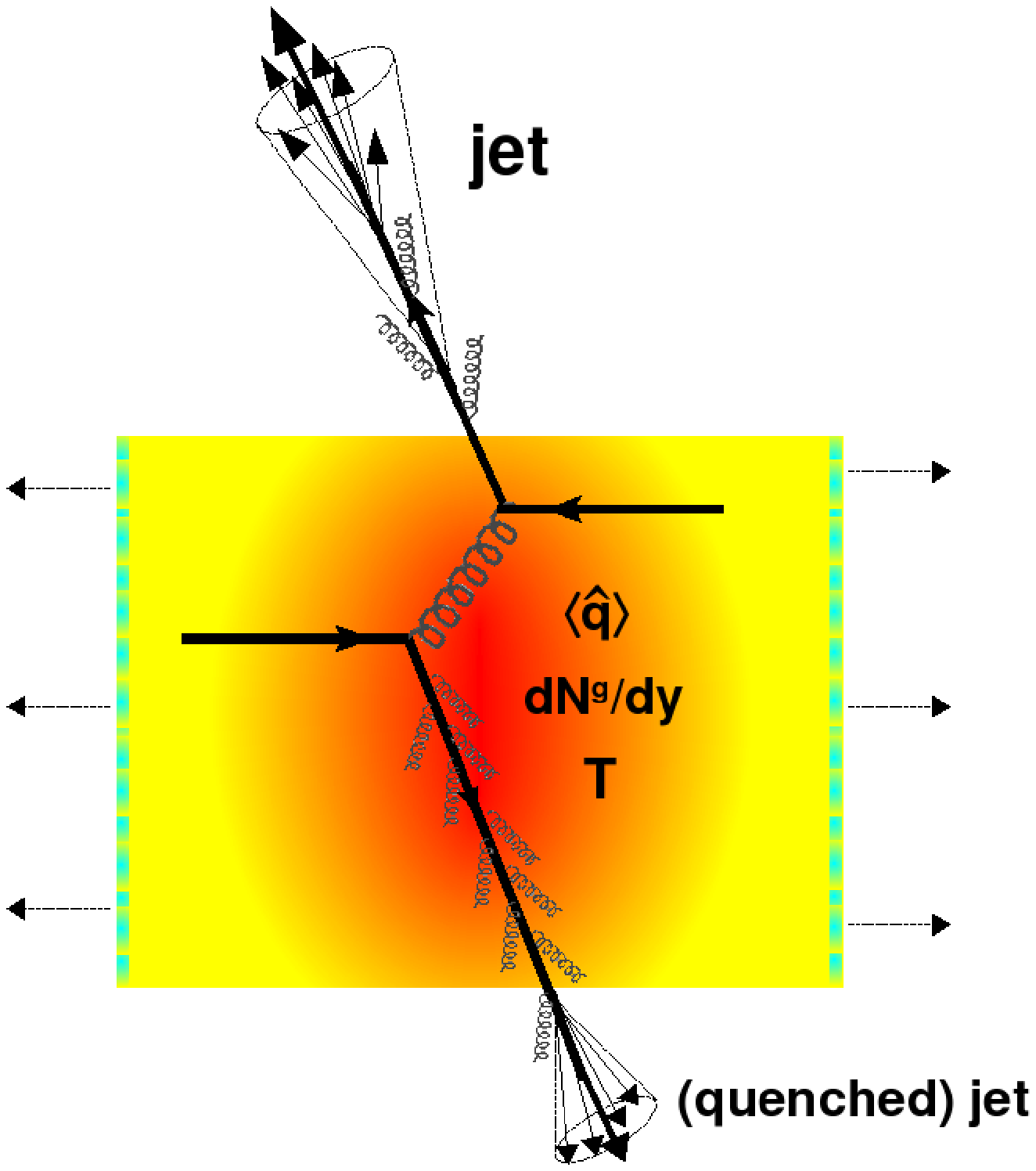}
\includegraphics[width=0.49\linewidth,height=6.2cm]{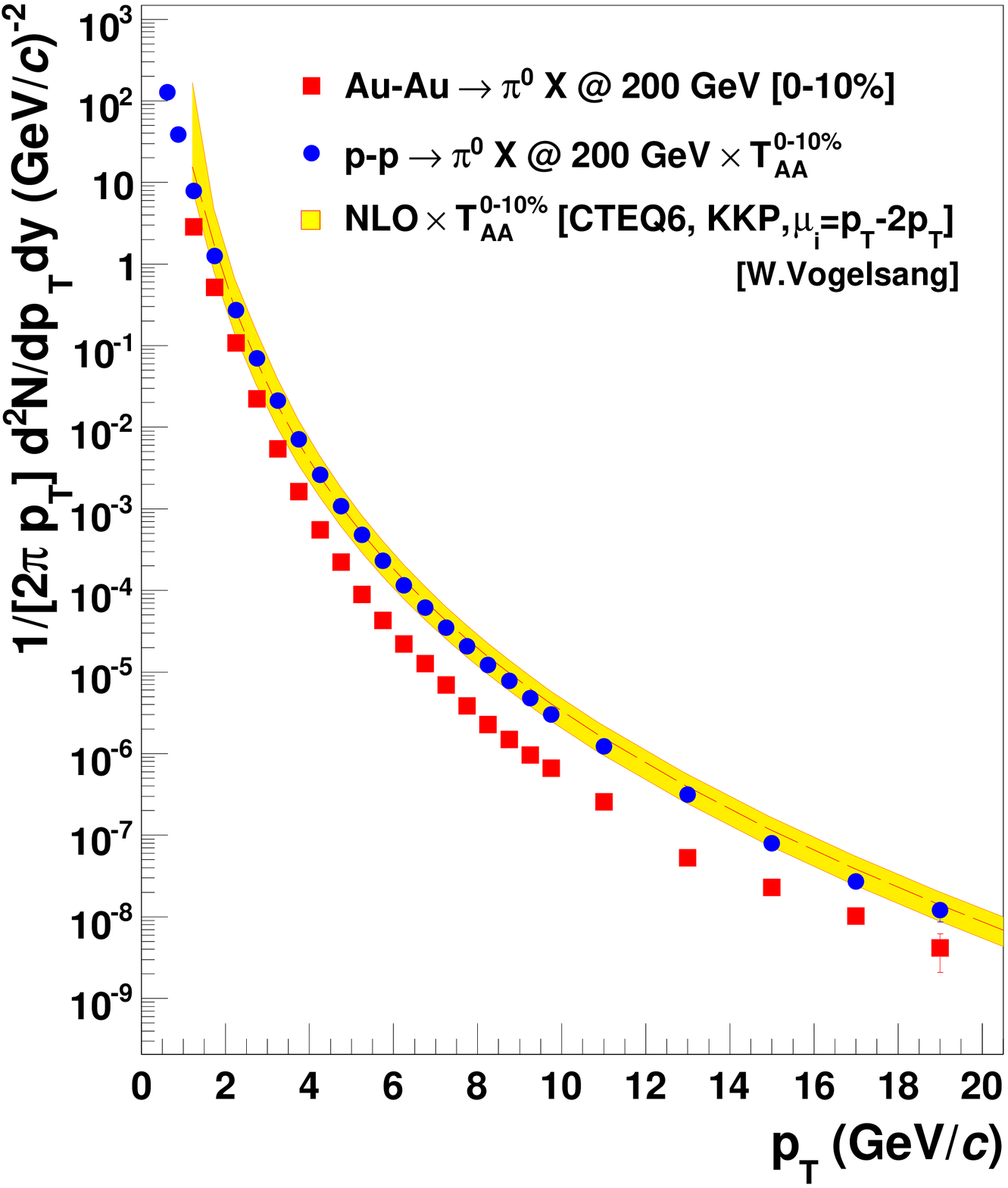}
\caption{Right: ``Jet quenching'' in a head-on heavy-ion collision: a fast parton traverses 
the dense plasma created (with transport coefficient $\qhat$, gluon density $dN^g/dy$ and temperature $T$),  
loses energy via ``gluonstrahlung'' and fragments into a (quenched) jet~\cite{d'Enterria:2009am}.
Left: Neutral pion spectrum measured by PHENIX at $\sqrtsnn$~=~200~GeV in central \AuAu\ (squares)~\protect\cite{Adare:2008qa}, 
compared to the ($T_{AA}$-scaled) spectrum in \pp\ collisions (circles)~\protect\cite{Adare:2007dg} 
and to a NLO pQCD calculation (yellow band)~\protect\cite{vogelsang_pi0}.}
\label{fig:jet_quench}
\end{figure}

One of the first proposed ``smoking guns'' of QGP formation was {\it jet quenching}~\cite{Bjorken:1982tu} 
i.e. the attenuation or disappearance of the spray of hadrons resulting from the fragmentation of a 
parton having suffered energy loss in the dense plasma formed in the reaction (Fig.~\ref{fig:jet_quench}, left).
The energy lost by a parton provides ``tomographic'' information of the matter properties
(temperature $T$, interaction coupling $\alpha$, thickness $L$): $\Delta E = f(E; T, \alpha, L)$~\cite{Peigne:2008wu}. 
The ``scattering power'' of the medium is often encoded in the {\it transport coefficient} which describes the 
average transverse momentum squared transferred to the traversing parton per unit path-length: 
$\qhat \equiv m_D^2/\lambda= \; m_D^2\; \rho \; \sigma$ (here $m_D$ is the medium Debye mass, $\rho$ its
density, and $\sigma$ the parton-matter interaction cross section)\footnote{For an equilibrated {\it gluon} 
plasma at $T=0.4$~GeV with coupling $\alpha_s\approx$~0.5 -- i.e. with density 
$\rho_g = 16/\pi^2 \;\zeta (3)\cdot T^3 \approx$ 15~fm$^{-3}$,  
Debye mass $m_D = (4 \pi \alpha_s)^{1/2} T \approx$~1~GeV, and cross section $\sigma_{gg}\approx$~1.5~mb --
one has $\qhat \simeq 2.2$~GeV$^2$/fm~\cite{Baier:2006fr}.}.
The dominant mechanism of energy loss of a fast parton in a dense QCD plasma is of radiative 
nature (``gluonstrahlung''): the parton loses energy mainly by medium-induced multiple gluon 
emission~\cite{gyulassy90,bdmps,glv,Wiedemann:2000za}. 
Jet quenching in \AaAa\ reactions is characterised by various observable 
consequences compared to the same ``QCD vacuum'' measurements in proton-proton (\pp) collisions:
(i) suppressed {\it high-$p_T$ hadron} spectrum ($dN_{AA}/dp_T$),  
(ii) unbalanced back-to-back {\it high-$p_T$ dihadron azimuthal correlations} ($dN_{pair}/d\phi$), 
and (iii) modified energy-particle flow (softer hadron spectra, larger multiplicity, increased 
angular broadening, ...) within the final {\it jets}.
A detailed review of these topics can be found in~\cite{d'Enterria:2009am}, 
of which a summary is given in the following sections. 


\section{High-$p_T$ single inclusive hadron production}
\label{sec:highpT_single}

\begin{figure}[htbp]
\centering
\includegraphics[width=0.49\linewidth,height=5.cm]{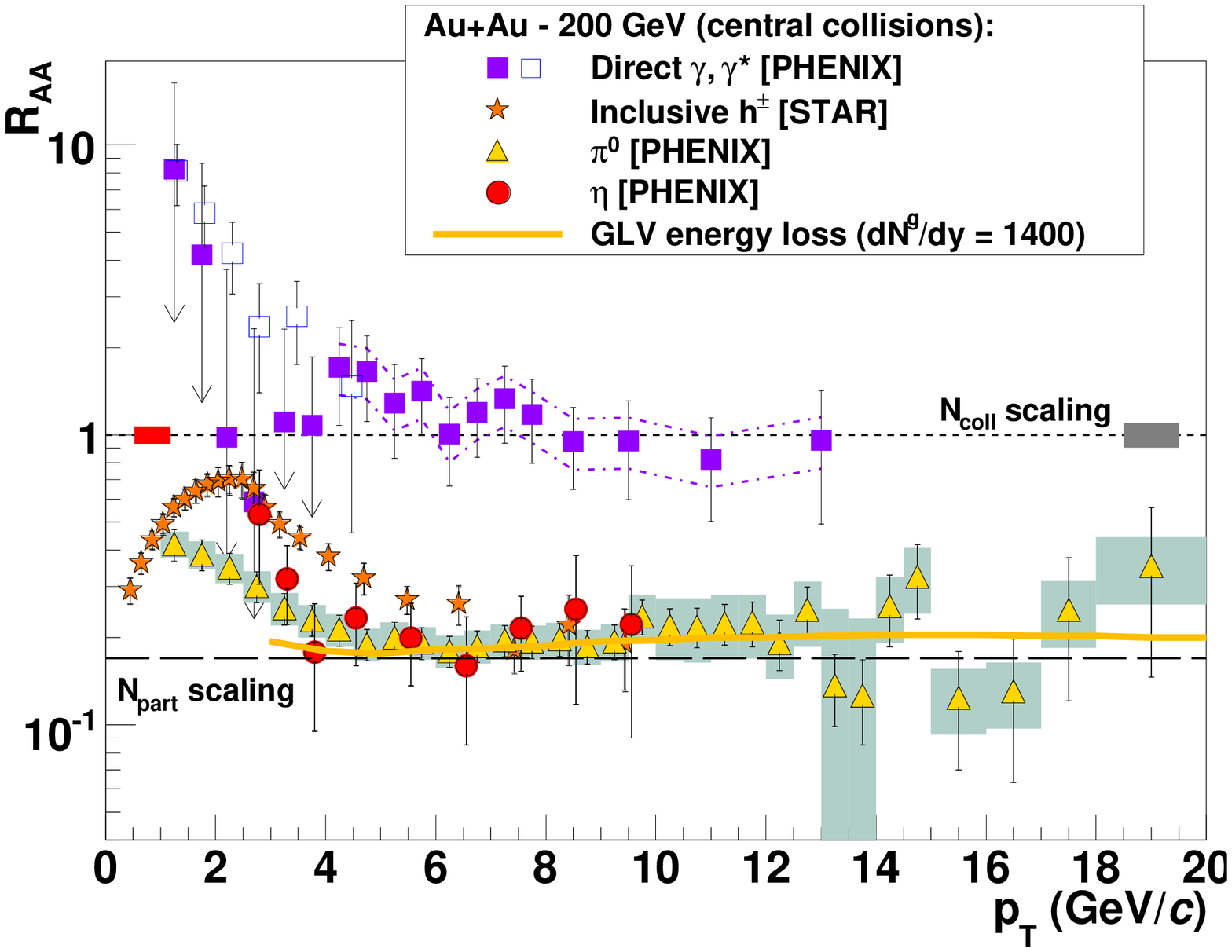}
\includegraphics[width=0.49\textwidth,height=5.cm]{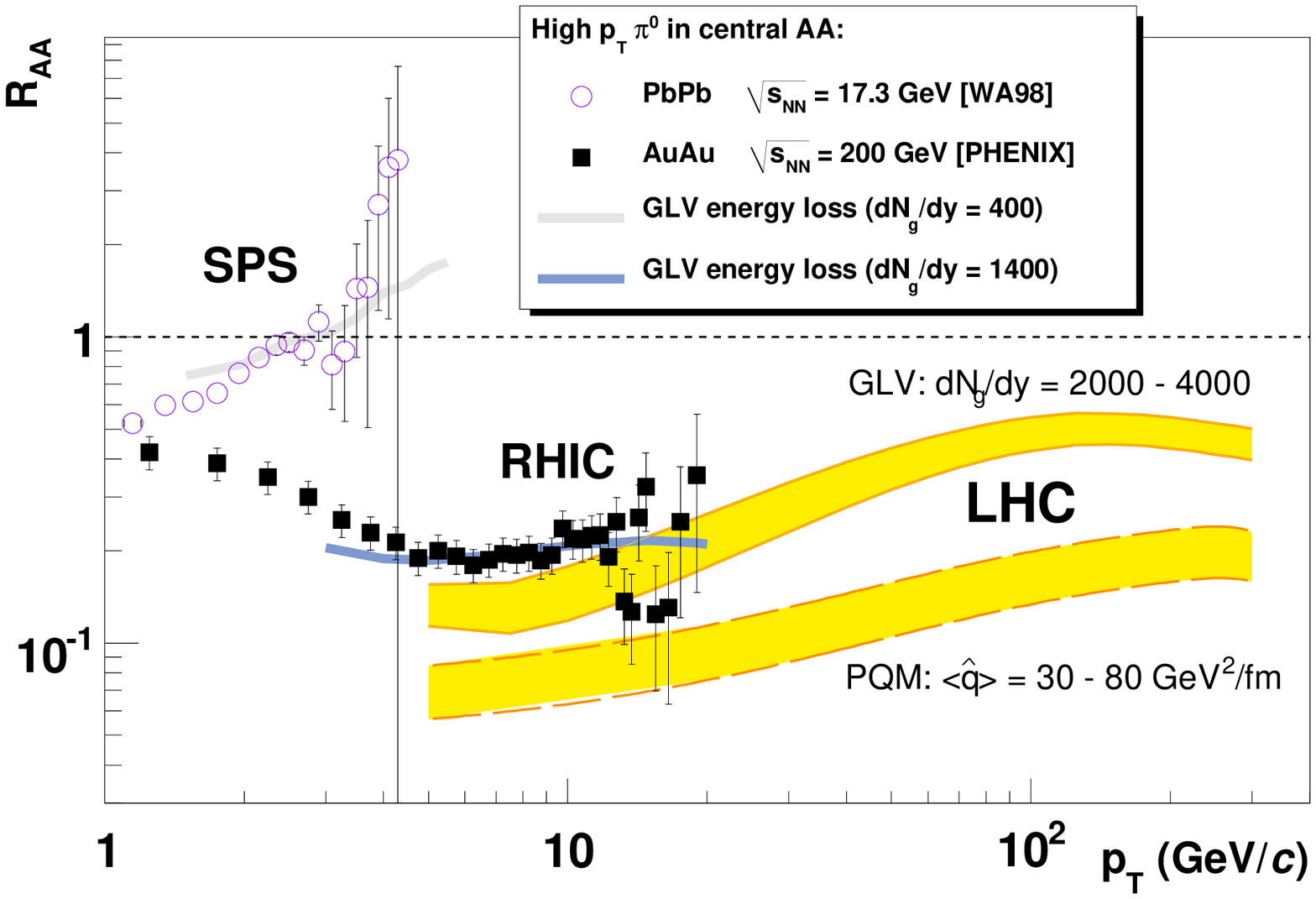}
\caption{Left: $R_{AA}(p_T)$ in central \AuAu\ at 200 GeV for $\pi^0$~\protect\cite{Adare:2008qa}, 
$\eta$~\protect\cite{phenix_hipt_pi0_eta_200}, charged hadrons~\protect\cite{star_hipt_200},
and direct $\gamma$~\protect\cite{phnx_gamma_AuAu200,Adare:2008fqa} compared to 
the GLV model ($dN^g/dy=$~1400, yellow curve)~\protect\cite{vitev_gyulassy}.
Right: $R_{AA}(p_{T})$ for $\pi^0$'s at SPS~\cite{Aggarwal:2001gn,d'Enterria:2004ig} 
and RHIC~\cite{Adare:2008qa} compared to  GLV calculations ($dN^g/dy$ = 400, 1400) and to 
predictions for central \PbPb\ at $\sqrtsnn$~=~5.5~TeV (yellow bands)~\protect\cite{Abreu:2007kv}: 
GLV ($dN^g/dy$ = 2000 -- 4000) and PQM $(\mqhat\approx$  30 -- 80 GeV$^2$/fm$)$.} 
\label{fig:RAAs}
\end{figure}

If a hard scattered parton suffers energy loss in a heavy-ion collision, the energy available for 
the hadrons issuing from its fragmentation will be reduced and 
their spectrum depleted compared to \pp\ collisions. The standard method to quantify the medium effects 
on the yield of a large-$p_T$ particle produced at rapidity $y$ in a \AaAa\ reaction is given by the 
{\it nuclear modification factor}:
\begin{equation}
\hspace{-0.8cm}
R_{AA}(p_{T},y;b)\,=\,\frac{d^2N_{AA}/dy dp_{T}}{\langle T_{AA}(b)\rangle\,\times\, d^2 \sigma_{pp}/dy dp_{T}}\;,
\mbox{ $T_{AA}(b)$ being the nuclear overlap function at $b$,}
\label{eq:R_AA}
\end{equation}
which measures the deviation of \AaAa\ at impact parameter $b$ from an 
incoherent superposition of nucleon-nucleon collisions ($R_{AA}$~=~1).
From the measured suppression factor one can determine various medium properties 
such as its transport parameter $\qhat$, via~$\langle\Delta E\rangle\propto\alpha_s\,\langle\hat{q}\rangle\,L^2$~\cite{bdmps,Wiedemann:2000za},
or its initial gluon density $dN^g/dy$, via
$\Delta E \propto \alpha_s^3\,C_R\,\frac{1}{A_\perp}\frac{dN^g}{dy}\,L$ (for an {\it expanding} plasma 
with {\it original} transverse area $A_\perp=\pi\,R_A^2\approx$~150~fm$^2$ and thickness $L$)~\cite{glv}.
We summarise the main high-$p_T$ hadroproduction results in \pp\ and \AaAa\ collisions, 
and confront them to jet quenching predictions.
\paragraph*{(a) Magnitude of the suppression and medium properties.}
Figure~\ref{fig:jet_quench} (right) shows the high-$p_T$ $\pi^0$ spectrum measured at $\sqrtsnn$~=~200 GeV
in central \AuAu~\cite{Adare:2008qa} compared to the \pp~\cite{Adare:2007dg} and NLO pQCD~\cite{vogelsang_pi0} 
spectra scaled by $T_{AA}$. The \AuAu\ data are suppressed by a factor of 4 -- 5 with respect to the \pp\ results. 
The corresponding $R_{AA}(p_T)$, Eq.~(\ref{eq:R_AA}), is shown in Fig.~\ref{fig:RAAs} (left).
Above $p_T\approx$~5~GeV/c, $\pi^0$~\cite{phenix_hipt_pi0_200}, $\eta$~\cite{phenix_hipt_pi0_eta_200}, 
and charged hadrons~\cite{star_hipt_200,phenix_hipt_200} show all a common factor of $\sim$5 suppression 
relative to the $R_{AA}$~=~1 expectation that holds for hard probes, such as direct 
photons~\cite{phnx_gamma_AuAu200,Adare:2008fqa}, which do no interact with the medium. 
The \AuAu\ high-$p_T$ suppression can be well reproduced by parton energy loss calculations
in a very dense system with initial gluon rapidity densities $dN^g/dy\approx$~1400 
(Gyulassy-L\'evai-Vitev curve in Fig.~\ref{fig:RAAs}, left)~\cite{vitev_gyulassy}, transport coefficients
$\qhat\approx$~13~GeV$^2$/fm~\cite{Adare:2008qa,pqm}, or plasma temperatures 
$T\approx$~0.4~GeV~\cite{Turbide:2005fk}. 
The consistency between the extracted $\qhat$, $dN^g/dy$ and $T$ values in the various models 
has been studied e.g. in~\cite{d'Enterria:2009am,Bass:2008rv}. 
Whereas the agreement between the fitted {\it thermodynamical} variables $dN^g/dy$ and $T$ is good, 
the values of the {\it transport} parameter $\qhat$ favoured by the data are 3 -- 4 times larger than 
perturbative estimates~\cite{Baier:2006fr}. An accord between the obtained $\qhat$ and $dN^g/dy$ 
can only be achieved assuming parton-medium cross-sections much larger than the $\sigma_{gg}\approx$~1.5~mb
LO expectation. Such an observation lends support to the {\it strongly-coupled} nature 
of the QGP produced at RHIC~\cite{gyulassy_mclerran04}.
\paragraph*{(b) Centre-of-mass energy dependence.}
As one increases the collision energy in nucleus-nucleus collisions, the produced plasma
reaches higher energy and particle densities, the system stays longer in the QGP phase, and 
correspondingly the traversing partons are more quenched. Figure~\ref{fig:RAAs} (right) 
compiles the measured $R_{AA}(p_T)$ for high-$p_T$ $\pi^0$ in central \AaAa\ 
collisions in the range $\sqrtsnn\approx$ 20 -- 200 GeV compared to parton energy loss 
calculations that assume the formation of a QGP with initial gluon densities 
in the range $dN^g/dy\approx$ 400 -- 1400~\cite{vitev_gyulassy,Vitev:2004gn} 
or, equivalently, averaged transport coefficients 
$\mean{\hat{q}}\approx$~3.5~--~13~GeV$^2$/fm~\cite{pqm}.
The theoretical predictions reproduce well the magnitude and shape of the experimental data.
The SPS $R_{AA}(p_T)$~\cite{Aggarwal:2001gn}, though consistent with unity~\cite{d'Enterria:2004ig}, 
is suppressed compared to the ``Cronin enhancement'' observed in peripheral \PbPb\ and $pPb$ collisions
at the same $\sqrtsnn$~\cite{Aggarwal:2007gw}.
\paragraph*{(c) $p_T$-dependence of the suppression.}
At RHIC top energies, the hadron quenching factor remains basically constant from 5~GeV/c up to the highest 
transverse momenta measured so far, $p_T\approx$~20~GeV/c (Fig.~\ref{fig:RAAs}).
Full calculations~\cite{vitev_gyulassy,pqm,Jeon:2003gi,Eskola:2004cr}
including the combined effect of (i) energy loss kinematics constraints, 
(ii) steeply falling $p_T$ spectrum of the scattered partons, and (iii) $\mathscr{O}($20\%) $p_T$-dependent 
(anti)shadowing differences between the proton and nuclear parton distribution functions (PDFs), result in an effectively flat $R_{AA}(p_T)$ 
as found in the data. The much larger kinematical range opened at LHC energies~\cite{Abreu:2007kv} 
will allow one to test the $p_T$-dependence of parton energy loss over a much wider domain than at RHIC
(yellow bands in Fig.~\ref{fig:RAAs}, right).
\paragraph*{(d) Centrality (system-size) dependence.}
The volume of the produced plasma in a heavy-ion collision can be ``dialed'' by modifying the 
overlap area between the colliding nuclei either by selecting a given impact-parameter $b$ -- 
i.e. by choosing more central or peripheral reactions -- or by colliding larger or smaller nuclei, 
e.g. Au~($A$~=~197) versus Cu~($A$~=~63).
The relative energy loss depends on the effective mass number $A_{\rm eff}$ or, equivalently, 
on the number of participant nucleons in the collision $N_{\rm part}$, as: 
${ \Delta E }/{E} \; \propto \; A_{\rm eff}^{2/3} \; \propto \; N_{\rm part}^{2/3}$~\cite{pqm,Vitev:2005he}.
The measured $R_{AA}(p_T)$ in {\it central} \CuCu\ at 22.4, 62.4, and 200\,GeV~\cite{Adare:2008cx} is 
a factor of $(A_{Au}/A_{Cu})^{2/3} \approx$~2 lower than in central \AuAu\ at the same energies. 
Yet, for a comparable $N_{part}$ value, the suppression in \AuAu\ and \CuCu\ is very similar. 
Fitting the $N_{part}$ dependence to $R_{AA} = (1 - \kappa\; N_{part}^{\alpha})^{n-2}$ 
yields $\alpha = 0.56 \pm 0.10$~\cite{Adare:2008qa}, consistent with parton energy loss calculations.
\paragraph*{(e) Path-length dependence.}
Experimentally, one can test the  dependence of parton suppression on the plasma thickness ($L$)
by exploiting the spatial asymmetry of the  system produced in {\it non-central} nuclear collisions.
Partons produced ``in plane'' (``out-of-plane'') i.e. along the short (long) direction of the ellipsoid 
matter with eccentricity $\epsilon$ will comparatively traverse a shorter (longer) thickness. 
PHENIX has measured the high-$p_T$ neutral pion suppression as a function 
of the angle with respect to the reaction plane, $R_{AA}(p_T,\phi)$~\cite{Adler:2006bw}.
Each azimuthal angle $\phi$ can be associated with an average medium path-length $L_{\epsilon}$ 
via a Glauber model. 
The energy loss is found to satisfy the expected $\Delta E \propto L$ 
dependence~\cite{glv} above a ``threshold'' length of $L\approx$~2~fm.
\begin{figure}[htbp]
\includegraphics[width=0.49\linewidth,height=5.2cm,clip]{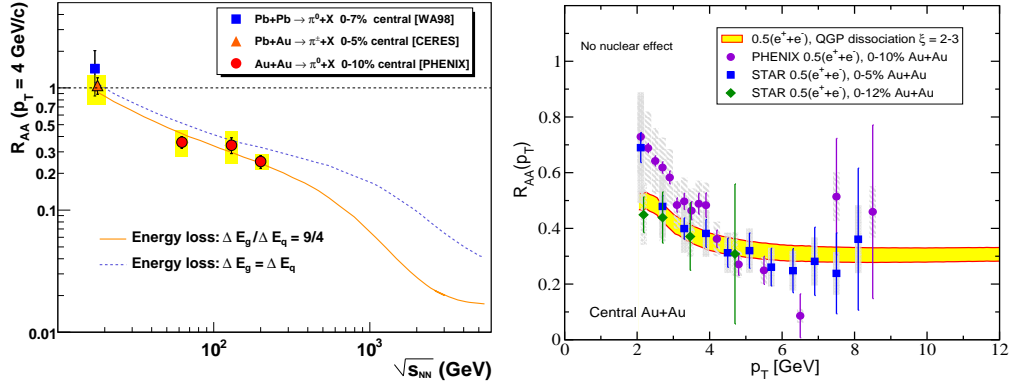}
\includegraphics[width=0.49\linewidth,height=5.cm,clip]{denterria_RAA_nonphot_elec_adilvitev.eps}
\caption{
Left: $R_{AA}(p_T$~=~4~GeV/c) for $\pi^0$  in central \AaAa\ collisions as function of collision energy
compared to non-Abelian (solid) and ``non-QCD'' (dotted) energy loss curves~\cite{dde_hp04,wang05}.
Right: $R_{AA}(p_T)$ for decay electrons from $D$ and $B$ mesons in central \AuAu\ at 
$\sqrt{s_{NN}}=200$~GeV~\cite{Adler:2005xv,Adare:2006nq,Abelev:2006db} 
compared to a model of $D$ and $B$ meson dissociation in the plasma~\cite{Adil:2006ra}.}
\label{fig:RAA_details}
\end{figure}
\paragraph*{(f) Non-Abelian (colour factor) dependence.}
The amount of energy lost by a parton in a medium is proportional to its colour Casimir factor: 
$C_A =$~3 for gluons, $C_F$~=~4/3 for quarks. Asymptotically, the probability 
for a gluon to radiate another gluon is $C_A/C_F$ = 9/4 times larger than for a quark and thus
$g$-jets are expected to be more quenched than $q$-jets in a QGP. One can test such a genuine 
{\it non-Abelian} property of QCD energy loss by measuring hadron suppression at a {\it fixed} 
$p_T$ for {\it increasing} $\sqrt{s}$~\cite{dde_hp04,wang05}.
At large (small) $x$, the PDFs are dominated by valence-quarks (``wee'' gluons) 
and consequently hadroproduction is dominated by quark (gluon) fragmentation. 
Figure~\ref{fig:RAA_details} (left) shows the $R_{AA}$ for 4-GeV/c pions measured 
at SPS and RHIC compared to two parton energy loss curves~\cite{wang05}.
The lower (upper) curve shows the expected $R_{AA}$ assuming a normal (arbitrary) behaviour 
with $\Delta E_g/\Delta E_q$~=~9/4 ($\Delta E_g = \Delta E_q$).
The experimental high-$p_T$ $\pi^0$ data supports\footnote{The use of high-$p_T$ (anti)protons 
(mostly coming from gluon fragmentation) as an alternative test of the colour charge 
dependence of the quenching~\cite{Mohanty:2008tw} is, unfortunately, distorted 
by the presence of extra non-perturbative mechanisms of baryon production (see discussion 
in~\cite{d'Enterria:2009am}).} the expected colour-factor dependence 
of $R_{AA}(\sqrtsnn)$~\cite{dde_hp04}.
\paragraph*{(g) Heavy-quark mass dependence.}
Due to the ``dead cone'' effect~\cite{deadcone}, the radiative energy loss for a charm (bottom) 
quark is 
a factor 1-$(m_Q/m_D)\approx$~25\% (75\%) smaller than for a light-quark. 
Yet, RHIC measurements~\cite{Adler:2005xv,Adare:2006nq,Abelev:2006db} 
of high-$p_T$ electrons from the semi-leptonic decays of $D$- and $B$-mesons 
(Fig.~\ref{fig:RAA_details}, right) indicate the same suppression for 
light and heavy mesons: $R_{AA}(Q)\sim R_{AA}(q,g)\approx$~0.2, 
in contradiction with parton energy loss predictions~\cite{djordj04,Armesto:2005mz}.
Various explanations have been proposed to solve the `heavy flavor puzzle'~\cite{d'Enterria:2009am}.
Among them is the observation that the hypothesis of {\it vacuum} hadronisation (after in-medium radiation) implicit in all 
formalisms may well not hold in the case of heavy quarks. The formation time of $D$- and $B$-mesons
is of order $\tau_{_{form}}\approx$~0.4 -- 1 fm/c respectively and, thus, one needs to account for 
the energy loss of the heavy-quark as well as the possible dissociation of the heavy-quark {\it meson} inside the QGP~\cite{Adil:2006ra}. 
The expected amount of suppression in that case is larger and consistent with the data (Fig.~\ref{fig:RAA_details}, right).


\section{High-$p_T$ di-hadron correlations}
\label{sec:}

\begin{figure}[htbp]
\includegraphics[width=0.49\linewidth,height=5.6cm,clip]{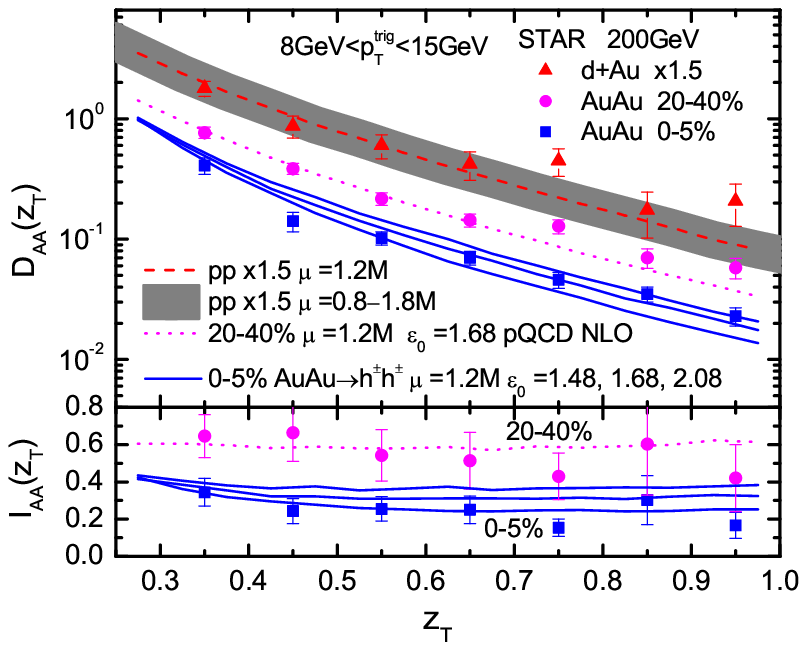}
\includegraphics[width=0.49\linewidth,height=5.8cm,clip]{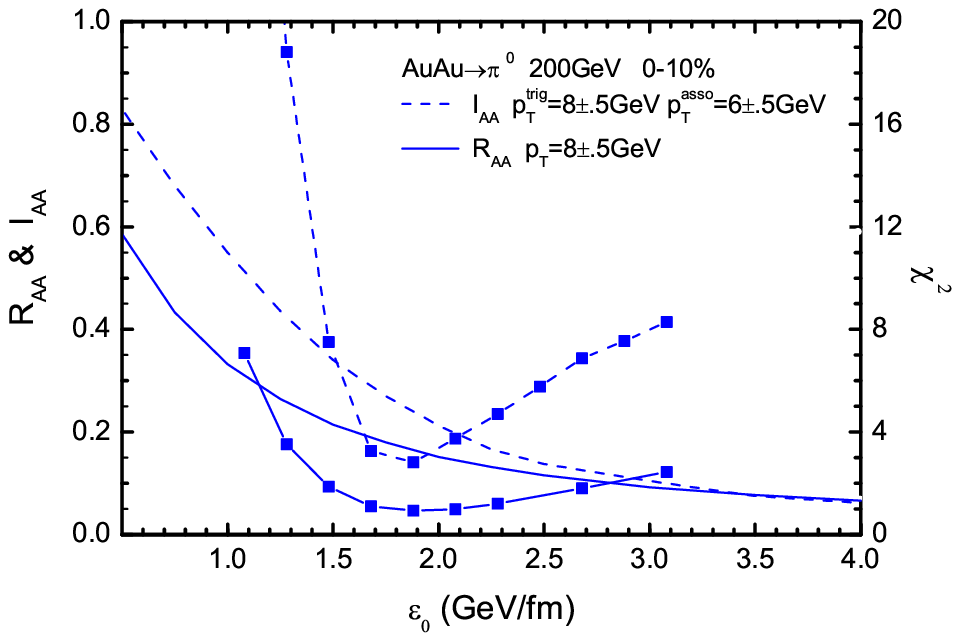}
\caption{Left: $D^{away}_{AA}(z_T)$ distributions for \dAu\ and \AuAu\ 
and $I_{AA}(z_T)$ ratio for central \AuAu\ at 200~GeV~\cite{star_jet_punchthrough}, 
compared to HT calculations for varying $\epsilon_0$ energy losses~\cite{Zhang:2007ja}. 
Right: Data vs. theory $\chi^2$ values for the fitted $\epsilon_0$ parameter~\cite{Zhang:2007ja}.}
\label{fig:IAA_qhat}
\end{figure}

Jet-like correlations in heavy-ion collisions can be measured on a {\it statistical} basis by selecting 
a high-$p_T$ {\it trigger} particle and measuring the azimuthal ($\Delta\phi = \phi - \phi_{trig}$) 
and pseudorapidity ($\Delta\eta = \eta - \eta_{trig}$) distributions of its {\it associated} hadrons 
($p_{T}^{assoc}<p_{T}^{trig}$):
$C(\Delta\phi,\Delta\eta) = \frac{1}{N_{trig}}\frac{d^2N_{pair}}{d\Delta\phi d\Delta\eta}$.
In \pp\ collisions, a dijet signal appears as two distinct back-to-back Gaussian-like peaks at $\Delta\phi\approx$ 0, $\Delta\eta\approx$ 0 
(near-side) and at $\Delta\phi\approx\pi$ (away-side). At variance with such a topology,
\begin{figure}[htbp]
\centering
\includegraphics[width=0.9\linewidth,height=4.cm]{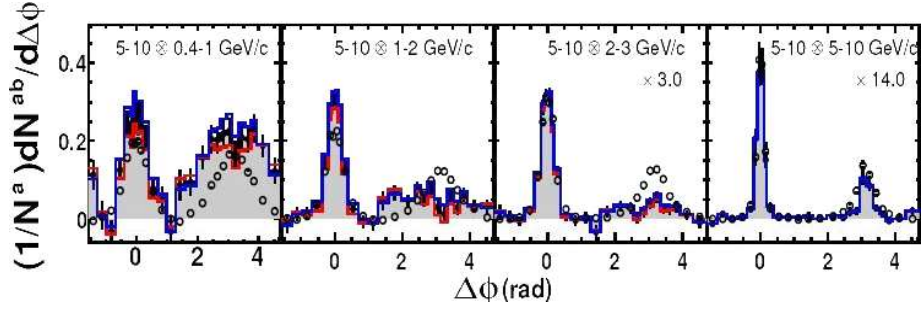}
\caption{Comparison of the azimuthal di-hadron correlation $dN_{pair}/d\Delta\phi d\eta$ for \pp\ (open symbols) 
and central \AuAu\ (histograms) at $\sqrtsnn$~=~200~GeV for $p_T^{trig}=$~5--10~GeV/c and increasingly
smaller (right to left) values of $p_T^{assoc}$~\protect\cite{Adare:2008cq}.}
\label{fig:dNdphi}
\end{figure}
Fig.~\ref{fig:dNdphi} shows the increasingly distorted back-to-back azimuthal correlations 
in high-$p_T$ triggered central \AuAu\ events as one decreases the $p_T$ of the associated hadrons
(right to left). Whereas, the \AuAu\ and \pp\ near-side peaks are similar for all $p_T$'s, 
the away-side peak is only present for the highest partner $p_T$'s but progressively disappears 
for less energetic partners~\cite{Adare:2007vu,Adare:2008cq}.
Early STAR results~\cite{star_hipt_awayside} showed a monojet-like topology with a complete 
disappearance of the opposite-side peak for $p_{T}^{assoc}\approx$~2~--~4~GeV/c.
The correlation strength over an azimuthal range $\Delta\phi$ between a trigger hadron $h_{t}$ and a partner hadron 
$h_{a}$ in the opposite azimuthal direction can be constructed as a function of the momentum fraction 
$z_T=p_{T}^{assoc}/p_{T}^{trig}$ via a ``pseudo-fragmentation function'':
\begin{equation}
D^{away}_{AA} (z_T) =
\int dp_{T}^{trig} \int dp_{T}^{assoc}
\int \limits_{\Delta\phi_{away}>130^o} d\Delta\phi\; \frac{d^3\sigma_{AA}^{h_t h_a}/d p_{T}^{trig}d p_{T}^{assoc}
d\Delta\phi}{d\sigma_{AA}^{h_t}/d p_{T}^{trig}}\,.
\label{eq:D_AA}
\end{equation}
shown in Fig.~\ref{fig:IAA_qhat} (top-left) 
compared to predictions of the HT model for various values of the $\epsilon_0$ 
parameter quantifying the amount of energy loss~\cite{Zhang:2007ja}.
Similarly to $R_{AA}$, the magnitude of the suppression of back-to-back jet-like two-particle 
correlations can be quantified with the ratio 
$\IAAa = D^{away}_{AA}/D^{away}_{pp}$. 
$\IAAa$ is found to decrease with increasing centrality, down to about 0.2~--~0.3 for the most 
central events (Fig.~\ref{fig:IAA_qhat}, bottom-left)~\cite{star_hipt_awayside,star_jet_punchthrough}. 
The right plot of Fig.~\ref{fig:IAA_qhat} shows the best $\epsilon_0\approx$~1.9~GeV/fm value that fits the 
measured $R_{AA}(p_T)$ and $I_{AA}(z_T)$ factors. Due to the irreducible presence of (unquenched) partons emitted 
from the surface of the plasma, the single-hadron quenching factor $R_{AA}(p_T)$ is in general 
less sensitive to the value of $\epsilon_0$ than the dihadron modification ratio $I_{AA}(z_T)$. 


\section{Jet observables}
\label{sec:}

\begin{figure}[htb]
\includegraphics[width=0.49\linewidth,height=5.2cm,clip]{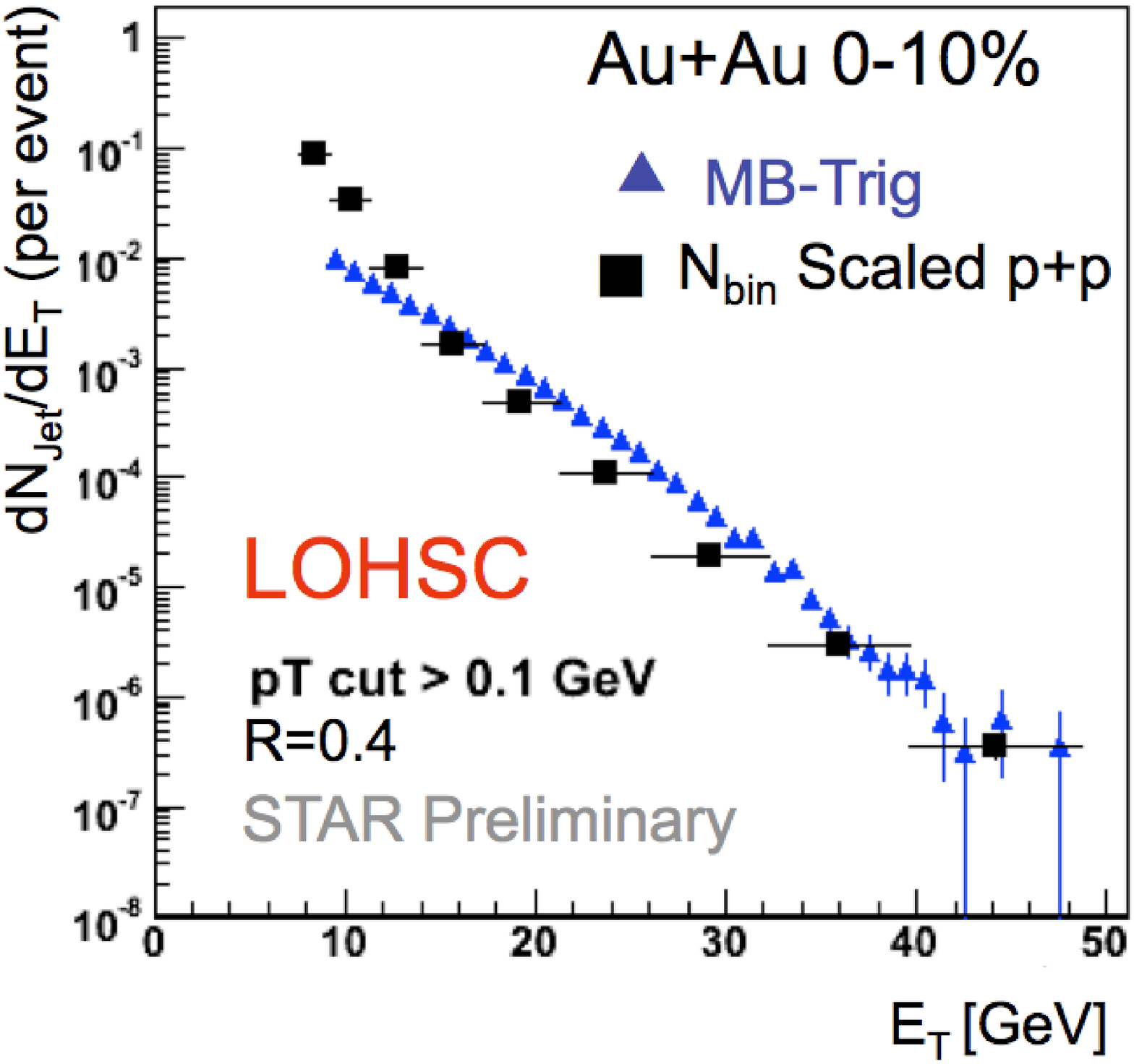}
\includegraphics[width=0.49\linewidth,height=5.2cm,clip]{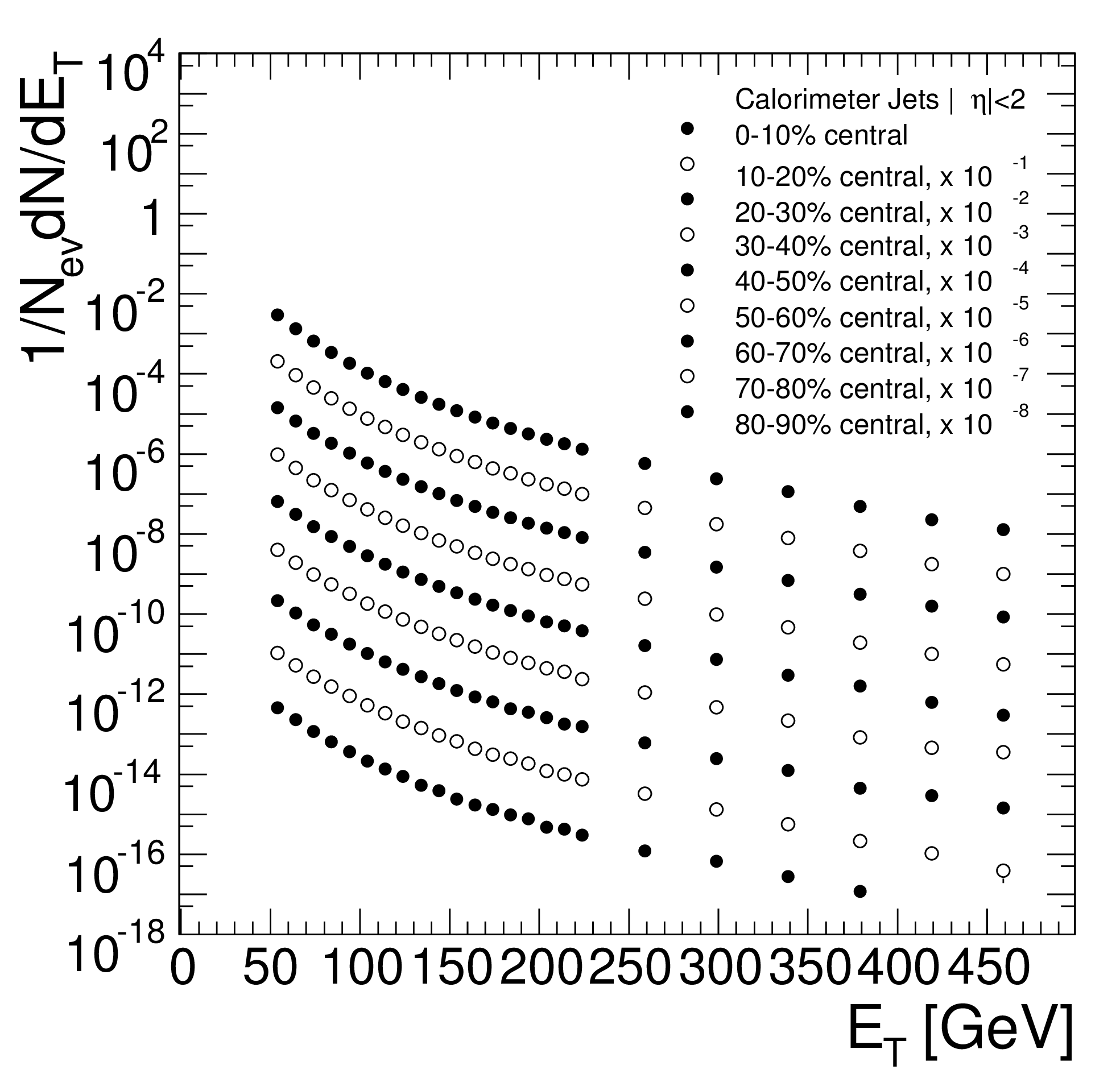}
\caption{Left: Preliminary STAR jet $E_T$ spectra in central \AuAu\ (triangles) and \pp\ (squares, scaled by $T_{AA}$) 
at 200~GeV~\cite{Putschke:2008wn}. Right: Jet spectra for various \PbPb\ centralities expected at 5.5~TeV 
in CMS ($\int\mathcal{L}\,dt$~=~0.5~nb$^{-1}$)~\cite{D'Enterria:2007xr}.}
\label{fig:jets}
\end{figure}

The measurements in \AaAa\ collisions of fully reconstructed (di)jets or of jets tagged by an 
away-side photon or $Z$-boson  allow one to investigate -- in much more detail than with single- 
or double- hadron observables -- the mechanisms of in-medium parton radiation as well as to 
characterise the matter properties through modified jet profiles~\cite{Salgado:2003rv,Vitev:2008rz} 
and fragmentation functions~\cite{Arleo:2008dn}. 
Experimental reconstruction of jets in nuclear reactions is an involved three-steps exercise
~\cite{d'Enterria:2009am}:
\begin{enumerate}
\item {\it Clustering algorithm}: 
The measured hadrons are clustered together, according to relative distances in momentum 
and/or space, following an {\it infrared- and collinear-safe} procedure which is also {\it fast} enough to be run 
over events with very high multiplicities. 
The $k_T$ and SISCone algorithms implemented 
in the \fastjet\ package~\cite{Cacciari:2005hq} fulfill all such conditions.
\item {\it Background subtraction}: Jets are produced on top of a large ``underlying event'' (UE) of hadrons 
coming from other (softer) parton-parton collisions in the same interaction. In central \PbPb\ collisions 
at the LHC one expects $E_T^{UE}\approx$~80~GeV (with large fluctuations) in a cone of radius 
$R = \sqrt{\Delta\eta^2+\Delta\phi^2}= 0.4$. 
Various UE subtraction techniques are available in combination with 
the $k_T$~\cite{Cacciari:2007fd}, UA1-cone~\cite{Blyth:2006vb} or iterative-cone~\cite{Kodolova:2007hd} algorithms.
\item {\it Jet energy corrections}: 
{\it Data-driven} methods are needed to experimentally control the {\it jet energy-scale} which 
is the single most important source of systematic uncertainties in the jet yield. The non-perturbative effects 
introduced by the UE and hadronisation can be gauged by comparing the sensitivity of the jet 
spectrum obtained with different Monte Carlo's~\cite{d'Enterria:2009am}.
\end{enumerate}
\vspace{0.2cm}
\noindent STAR~\cite{Putschke:2008wn} has a preliminary measurement of jets in \AuAu\ at 200~GeV 
(Fig.~\ref{fig:jets}, left) using a cone algorithm with $R$~=~0.4, and estimating the UE
background from the average energy in cones {\it without} seeds. Control of the jet energy corrections 
is still work in progress. Jet physics will definitely benefit from the highest energies (and therefore
statistics) available at the LHC. The expected $p_T$ reach in \PbPb\ at 5.5~TeV is as large as 
$p_T\approx$~500~GeV/c (Fig.~\ref{fig:jets}, right). 

\begin{figure}[htbp]
\includegraphics[width=0.49\linewidth,height=4.5cm,clip]{denterria_borghiniwiedemann_humpedback.eps}\hspace{0.25cm}
\includegraphics[width=0.49\linewidth,height=4.5cm,clip]{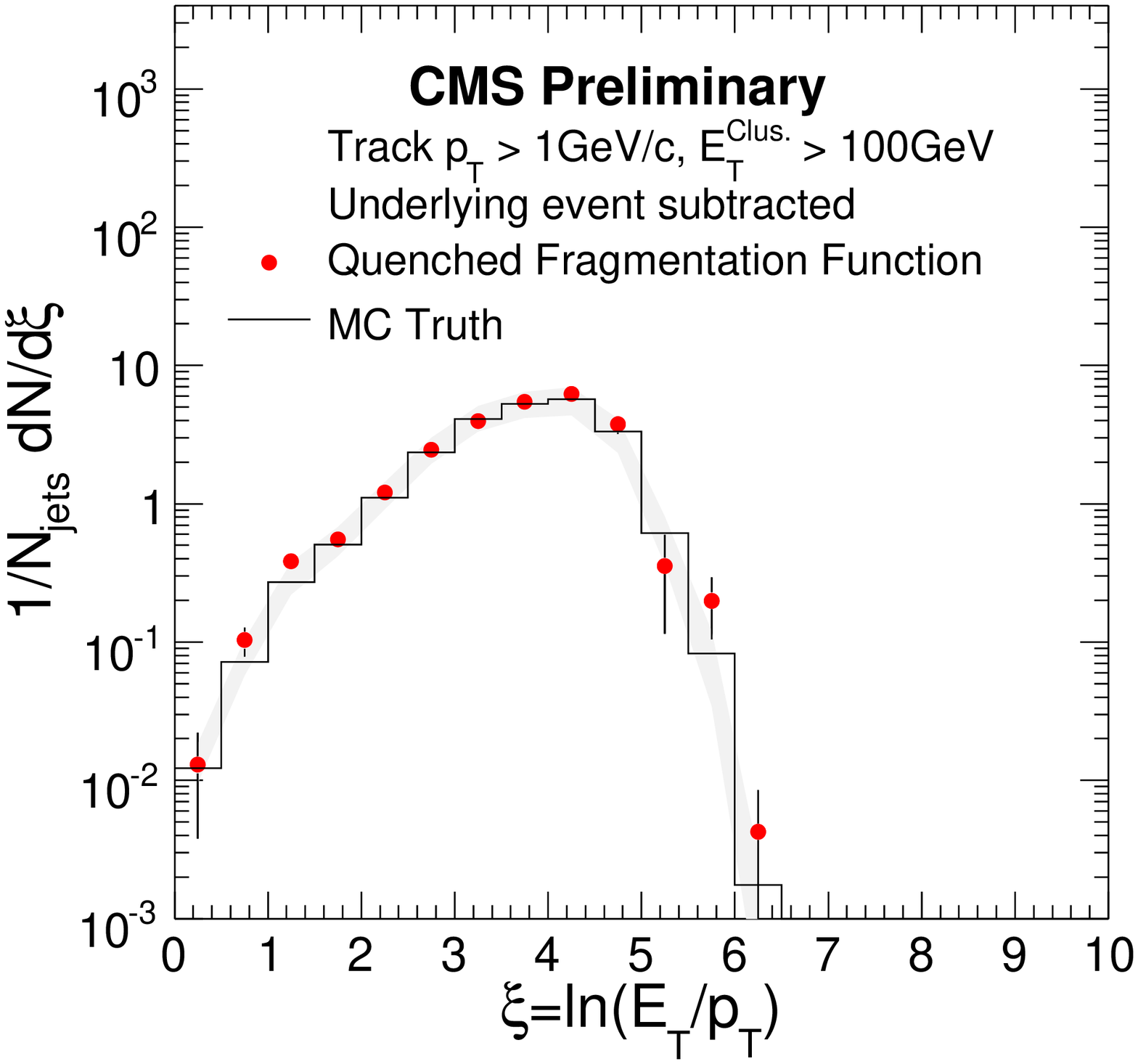}
\caption{Left: Single inclusive distribution of hadrons vs. $\xi=\ln\,(E_{jet}/p)$ for a 17.5-GeV jet in $e^+e^-$ 
collisions (TASSO data) compared to QCD radiation predictions in the vacuum (solid curve) and in-medium 
(dashed curve)~\cite{Borghini:2005em}. 
Right: FFs as a function of  $\xi$ for quenched partons obtained in CMS $\gamma$-jet simulations 
for central Pb-Pb at 5.5 TeV (0.5~nb$^{-1}$)~\protect\cite{Loizides:2008pb}.}
\label{fig:mFFs}
\end{figure}

The $\gamma$-jet (and $Z$-jet) channel provides a very clean means to determine medium-modified parton fragmentation
functions (FFs)~\cite{Wang:1996yh,Arleo:2004xj}. Since the prompt $\gamma$ is not affected by final-state interactions, 
its transverse energy ($\etg$) can be used as a proxy of the away-side parton energy ($\etj\approx\etg$) {\it before} 
any jet quenching. The FF, defined as the normalised distribution of hadron momenta $1/{N_{jets}}\,dN/dz$ relative to 
that of the parent parton $\etj$, can then be constructed using $\z = p_T/\etj$ or, similarly, $\xi=\log(\etj/p_T)=-\log(z)$, 
for all particles with momentum $p_T$ associated with a jet. 
In a QCD medium, energy loss shifts parton energy from high-$z$ to low-$z$ hadrons~\cite{Borghini:2005em}, resulting in a 
higher ``hump-back plateau'' in the FFs at intermediate $\xi\approx$~3~--~4 values (Fig.~\ref{fig:mFFs}, left).
Full simulation studies of the $\gamma$-jet channel in central Pb-Pb (Fig.~\ref{fig:mFFs}, right)
indicate that medium modified FFs are measurable with small uncertainties in the ranges 
$z<$~0.7 and 0.2~$<\xi<$~6~\cite{Loizides:2008pb}.

\section{Summary}
\label{sec:}

The analysis of jet structure modifications in heavy-ion collisions provides 
quantitative ``tomographic'' information on the thermodynamical and transport properties 
of the strongly interacting medium produced in the reactions.
At RHIC energies (up to $\sqrtsnn$~=~200~GeV), strong suppression of the yields of 
high-$p_T$ single hadrons and of dihadron azimuthal correlations, 
have been observed in central \AuAu\ collisions.
Most of the properties of the observed 
suppression 
are in quantitative agreement with the predictions of parton energy loss models in a very dense QCD plasma.
The confrontation of these models to the data permits to derive the initial 
gluon density $dN^g/dy\approx$ 1400 and transport coefficient 
$\hat{q} =\mathscr{O}($10~GeV$^2$/fm$)$ of the produced medium at RHIC.
At the upcoming LHC energies, the detailed analysis of jet spectra, jet shapes and the 
extraction of medium-modified parton-to-hadron fragmentation functions 
promise to fully unravel the mechanisms of parton energy loss 
in QCD matter. The study of jet quenching phenomena proves an excellent tool 
to expand our knowledge of the dynamics of the strong interaction at extreme conditions 
of temperature and density.

\section*{Acknowledgments}

\noindent
Special thanks to Itzhak~Tserruya and the organisers of PANIC'08 for their kind invitation.
Support by the 6th EU Framework Programme contract MEIF-CT-2005-025073 is acknowledged.

%

%

\begin{thebibliography}{00}

\def\PLB{{Phys. Lett.}~{\bf B}}
\def\PLC{Phys. Repts.\ }
\def\PRL{Phys. Rev. Lett.\ }
\def\PRD{{Phys. Rev.}~{\bf D}}
\def\PRC{{Phys. Rev.}~{\bf C}}
\def\NPA{{Nucl. Phys.}~{\bf A}}
\def\NPB{{Nucl. Phys.}~{\bf B}}
\def\EPJ{{Eur. Phys. J.}~{\bf C}}
\def\JPG{{J. Phys}~{\bf G}}
\def\JHEP{{J. High Energy Phys.}~}

\bibitem{d'Enterria:2006su}
D.~d'Enterria,
  J.\ Phys.\ G {\bf 34}, S53 (2007)

\bibitem{latt}
  M.~Cheng {\it et al.},
  Phys.\ Rev.\  D {\bf 74}, 054507 (2006);
  Y.~Aoki {\it et al.}, 
  Phys.\ Lett.\  B {\bf 643}, 46 (2006)

\bibitem{Bjorken:1982tu}
  J.~D.~Bjorken,
  FERMILAB-PUB-82-059-THY (1982)

\bibitem{d'Enterria:2009am}
  D.~d'Enterria,
  arXiv:0902.2011 [nucl-ex]

\bibitem{Adare:2008qa}
  A.~Adare {\it et al.}  [PHENIX Collaboration],
 Phys.\ Rev.\ Lett.\  {\bf 101}, 232301 (2008)

\bibitem{Adare:2007dg}
  A.~Adare {\it et al.}  [PHENIX Collaboration],
  Phys.\ Rev.\  D {\bf 76}, 051106 (2007)

\bibitem{vogelsang_pi0}
F.~Aversa, P.~Chiappetta, M.~Greco and J.~P.~Guillet,
  Nucl.\ Phys.\  B {\bf 327}, 105 (1989);
B.~Jager, A.~Schafer, M.~Stratmann and W.~Vogelsang,
  Phys.\ Rev.\  D {\bf 67}, 054005 (2003);
W.~Vogelsang (private communication)

\bibitem{Peigne:2008wu}
  S.~Peign\'e and A.~V.~Smilga,
  arXiv:0810.5702 [hep-ph]

\bibitem{Baier:2006fr}
  R.~Baier and D.~Schiff,
  JHEP {\bf 0609}, 059 (2006)

\bibitem{gyulassy90}
M.~Gyulassy, M.~Pl\"umer, \PLB{\bf 243}, 432 (1990); 
X.N.~Wang, M.~Gyulassy, \PRL {\bf 68}, 1480 (1992) 

\bibitem{bdmps}R.~Baier, Y.L.~Dokshitzer, A.H.~Mueller, S.~Peign\'e and D.~Schiff, \NPB{\bf 484}, 265 (1997);
R.~Baier, D.~Schiff, B.G.~Zakharov, Ann.\ Rev.\ Nucl.\ Part.\ Sci.\ {\bf 50}, 37 (2000)

\bibitem{glv}M.~Gyulassy, P.~Levai and I.~Vitev, \PRL {\bf 85}, 5535 (2000); 
\NPB {\bf 594}, 371 (2001) 

\bibitem{Wiedemann:2000za}
  U.~A.~Wiedemann,
  Nucl.\ Phys.\  B {\bf 588}, 303 (2000)

\bibitem{phenix_hipt_pi0_200}S.S.~Adler {\it et al.} [PHENIX Collaboration], \PRL {\bf 91}, 072301 (2003)

\bibitem{phenix_hipt_pi0_eta_200}
S.~S.~Adler {\it et al.}  [PHENIX Collaboration],
  Phys.\ Rev.\ Lett.\  {\bf 96}, 202301 (2006)

\bibitem{star_hipt_200}
J.~Adams {\it et al.} [STAR Collaboration], \PRL {\bf 91}, 172302 (2003)

\bibitem{phenix_hipt_200}S.~S.~Adler {\it et al.} [PHENIX Collaboration], \PRC {\bf 69}, 034910 (2004)

\bibitem{phnx_gamma_AuAu200}
S.~S.~Adler {\it et al.}  [PHENIX Collaboration],
  Phys.\ Rev.\ Lett.\  {\bf 94}, 232301 (2005)

\bibitem{Adare:2008fqa}
  A.~Adare {\it et al.}  [PHENIX Collaboration],
  arXiv:0804.4168 [nucl-ex]

\bibitem{vitev_gyulassy}I.~Vitev and M.~Gyulassy, \PRL {\bf 89}, 252301 (2002); 
I.~Vitev, J.\ Phys.\ G {\bf 30}, S791 (2004) 

\bibitem{Aggarwal:2001gn}
M.~M.~Aggarwal {\it et al.}  [WA98 Collaboration],
  Eur.\ Phys.\ J.\  C {\bf 23}, 225 (2002)

\bibitem{d'Enterria:2004ig}
  D.~d'Enterria,
  Phys.\ Lett.\  B {\bf 596}, 32 (2004)

\bibitem{Abreu:2007kv}
  N.~Armesto {\it et al.},
  J.\ Phys.\ G {\bf 35}, 054001 (2008)

\bibitem{pqm}
  A.~Dainese, C.~Loizides and G.~Paic,
  Eur.\ Phys.\ J.\  C {\bf 38}, 461 (2005)

\bibitem{Turbide:2005fk}
  S.~Turbide, C.~Gale, S.~Jeon and G.~D.~Moore,
  Phys.\ Rev.\  C {\bf 72}, 014906 (2005)

\bibitem{Bass:2008rv}
  S.~A.~Bass {\it et al.}, 
  arXiv:0808.0908 [nucl-th]

\bibitem{gyulassy_mclerran04}
M.~Gyulassy and L.~McLerran,
  Nucl.\ Phys.\  A {\bf 750}, 30 (2005)

\bibitem{Vitev:2004gn}
  I.~Vitev,
  Phys.\ Lett.\  B {\bf 606}, 303 (2005)

\bibitem{Aggarwal:2007gw}
  M.~M.~Aggarwal {\it et al.}  [WA98 Collaboration],
  Phys.\ Rev.\ Lett.\  {\bf 100}, 242301 (2008)


\bibitem{Jeon:2003gi}
  S.~Jeon and G.~D.~Moore,
  Phys.\ Rev.\  C {\bf 71}, 034901 (2005)

\bibitem{Eskola:2004cr}
  K.~Eskola, H.~Honkanen, C.~Salgado and U.~Wiedemann,
  Nucl.\ Phys.\  A {\bf 747}, 511 (2005)

\bibitem{Vitev:2005he}
  I.~Vitev,
  Phys.\ Lett.\  B {\bf 639}, 38 (2006)

\bibitem{Adare:2008cx}
  A.~Adare {\it et al.}  [PHENIX Collaboration],
  Phys.\ Rev.\ Lett.\  {\bf 101}, 162301 (2008)

\bibitem{Adler:2006bw}
  S.~S.~Adler {\it et al.}  [PHENIX Collaboration],
  Phys.\ Rev.\  C {\bf 76}, 034904 (2007)

\bibitem{dde_hp04}D.~d'Enterria, \EPJ {\bf 43}, 295 (2005)

\bibitem{wang05}Q.~Wang and X.N.~Wang, \PRC {\bf 71}, 014903 (2005)

\bibitem{Adler:2005xv}
  S.~S.~Adler {\it et al.}  [PHENIX Collaboration],
  Phys.\ Rev.\ Lett.\  {\bf 96}, 032301 (2006)

\bibitem{Adare:2006nq}
  A.~Adare {\it et al.}  [PHENIX Collaboration],
  Phys.\ Rev.\ Lett.\  {\bf 98}, 172301 (2007)

\bibitem{Abelev:2006db}
  B.~I.~Abelev {\it et al.}  [STAR Collaboration],
  Phys.\ Rev.\ Lett.\  {\bf 98}, 192301 (2007)


\bibitem{Adil:2006ra}
  A.~Adil and I.~Vitev,
  Phys.\ Lett.\  B {\bf 649}, 139 (2007)

\bibitem{Mohanty:2008tw}
  B.~Mohanty  [STAR Collaboration],
  J.\ Phys.\ G {\bf 35}, 104006 (2008)

\bibitem{deadcone}
  Y.~L.~Dokshitzer and D.~E.~Kharzeev,
  Phys.\ Lett.\  B {\bf 519}, 199 (2001)

\bibitem{djordj04}M.~Djordjevic, M.~Gyulassy and S.~Wicks, \PRL {\bf 94}, 112301 (2005)

\bibitem{Armesto:2005mz}
  N.~Armesto, M.~Cacciari, A.~Dainese, C.~A.~Salgado and U.~A.~Wiedemann,
  Phys.\ Lett.\  B {\bf 637}, 362 (2006)

\bibitem{Adare:2008cq}
  A.~Adare {\it et al.}  [PHENIX Collaboration],
  Phys.\ Rev.\  C {\bf 78}, 014901 (2008)

\bibitem{Adare:2007vu}
  A.~Adare {\it et al.}  [PHENIX Collaboration],
  Phys.\ Rev.\  C {\bf 77}, 011901 (2008)

\bibitem{star_hipt_awayside}C.~Adler {\it et al.} [STAR Collaboration], \PRL {\bf 90}, 082302 (2003)


\bibitem{Zhang:2007ja}
  H.~Zhang, J.~F.~Owens, E.~Wang and X.~N.~Wang,
  Phys.\ Rev.\ Lett.\  {\bf 98}, 212301 (2007)

\bibitem{star_jet_punchthrough}
  J.~Adams {\it et al.}  [STAR Collaboration],
  Phys.\ Rev.\ Lett.\  {\bf 97}, 162301 (2006)

\bibitem{Salgado:2003rv}
  C.~A.~Salgado and U.~A.~Wiedemann,
  Phys.\ Rev.\ Lett.\  {\bf 93}, 042301 (2004)

\bibitem{Vitev:2008rz}
  I.~Vitev, S.~Wicks and B.~W.~Zhang,
  arXiv:0810.2807 [hep-ph]

\bibitem{Arleo:2008dn}
  F.~Arleo,
  arXiv:0810.1193 [hep-ph]

\bibitem{Cacciari:2005hq}
  M.~Cacciari and G.~P.~Salam,
  Phys.\ Lett.\  B {\bf 641}, 57 (2006)

\bibitem{Cacciari:2007fd}
  M.~Cacciari and G.~P.~Salam,
  Phys.\ Lett.\  B {\bf 659}, 119 (2008)

\bibitem{Blyth:2006vb}
  S.~L.~Blyth {\it et al.},
  J.\ Phys.\ G {\bf 34}, 271 (2007)

\bibitem{Kodolova:2007hd}
  O.~Kodolova, I.~Vardanian, A.~Nikitenko and A.~Oulianov,
  Eur.\ Phys.\ J.\  C {\bf 50}, 117 (2007)

\bibitem{Putschke:2008wn}
  J.~Putschke  [STAR Collaboration],
arXiv:0809.1419; 
  S.~Salur  [STAR Collaboration],
  arXiv:0810.0500 [nucl-ex]

\bibitem{D'Enterria:2007xr}
  D.~d'Enterria (ed.) {\it et al.}  [CMS Collaboration],
  J.\ Phys.\ G {\bf 34}, 2307 (2007)

\bibitem{Wang:1996yh}
  X.~N.~Wang, Z.~Huang and I.~Sarcevic,
  Phys.\ Rev.\ Lett.\  {\bf 77}, 231 (1996)

\bibitem{Arleo:2004xj}
  F.~Arleo, P.~Aurenche, Z.~Belghobsi and J.~P.~Guillet,
  JHEP {\bf 0411}, 009 (2004) 

\bibitem{Borghini:2005em}
  N.~Borghini and U.~A.~Wiedemann,
  arXiv:hep-ph/0506218

\bibitem{Loizides:2008pb}
  C.~Loizides  [CMS Collaboration],
  J.\ Phys.\ G {\bf 35}, 104166 (2008)

\end{thebibliography}
\end{document}